\documentclass{JHEP3}
\def\ba{\begin{array}}
\def\ea{\end{array}}
\def\be{\begin{equation}}
\def\ee{\end{equation}}
\def\lbl{\label}

\def\x0{\x_0}
\def\x1{\x_1}

\def\real{ {\bf R} }
\def\compl{{\bf C}}
\def\cd{{\cal {D}}}
\def\cg{{\cal {G}}}
\def\tcg{{\tilde{\cal G}}}
\def\lie{{\cal {L}}}

\title{On modular spaces of semisimple Drinfeld doubles}

\author{Libor \v Snobl 
\\  Faculty of Nuclear Sciences and Physical Engineering,
\\  Czech Technical University, 
\\  B\v rehov\'a 7, 115 19 Prague 1, Czech Republic \\
\email{Libor.Snobl@fjfi.cvut.cz}}
\abstract{We construct modular spaces of all 6--dimensional real semisimple Drinfeld doubles, 
i.e. the sets of all possible decompositions of the Lie algebra of the Drinfeld double into Manin triples.
These modular spaces are significantly different from the known one for Abelian Drinfeld double, since 
some of these Drinfeld doubles allow decomposition into several non--isomorphic Manin triples and their 
modular spaces are therefore written as unions of homogeneous spaces of different dimension.
Implications for Poisson--Lie T--duality and especially Poisson--Lie T--plurality are mentioned.
} 

\begin{document}

\section{Introduction}
The interest in dualities of field theories, especially  the string theory, in the middle 90s has led to investigation of duality of 
$\sigma$--models and in this context to the discovery of 
a generalization of Abelian \cite{buscher1},\cite{buscher2} and non--Abelian T--duality \cite{ossaquevedo}, the so--called 
Poisson--Lie T--duality \cite{klse:dna},\cite{kli:pltd}. A crucial role in these considerations plays the notion of Manin triple and Drinfeld double.

Since then, several non--trivial examples of Poisson--Lie T--dual models, i.e. not contained in the class of non--Abelian T--dual models, were constructed and considered 
both on the classical and quantum level, e.g. the pair of models on one of the Manin triples of the Drinfeld double $SO(3,1)$, see \cite{vall:su2},\cite{sfe:pltd}. 

Moreover  also the knowledge of the modular spaces of Drinfeld doubles, i.e. the complete sets of their decompositions 
into different Manin triples, is of interest. Such modular spaces can be interpreted as spaces of $\sigma$--models mutually 
connected by Poisson--Lie T--duality. Consequently, if one is able to solve one of the $\sigma$--models, the solutions of 
all other models in the modular space follow. The knowledge of modular spaces also enables to investigate the structure 
of equivalent D--brane and open string theories on subgroups of the given Drinfeld double, see \cite{klse:opdb}.
Up to now only rather trivial examples of modular spaces are known. 

In the present paper we derive several new examples of modular spaces of Drinfeld doubles, rather distinct from the known ones. 
This will enable  to investigate in future concrete examples of the general structures and procedures
suggested in literature, e.g. \cite{klse:dna},\cite{kli:pltd},\cite{klse:opdb}, to gain better insight and understanding of them and possibly 
to discover some of their still unknown properties.
Firstly we recall the definitions of Drinfeld double and Manin triple and define the modular space of Drinfeld double. Secondly we give 
some basic information about Poisson--Lie T--duality and its connection with the algebraic structures just defined. Then we present a construction of 
the modular spaces of six--dimensional semisimple Drinfeld doubles on the groups 
$SL(2,\real) \times SL(2,\real)$ and $SO(1,3)_{+}$. 

\section{Manin triples, Drinfeld doubles and their modular spaces}

The {\bf Drinfeld double} $D$ is defined as a connected Lie group such that its Lie algebra 
$\cd$ equipped by a symmetric ad--invariant nondegenerate bilinear form 
$\langle .,.\rangle $ can be decomposed into a pair of 
subalgebras $\cg$, $\tcg$ maximal isotropic with respect to $\langle .,.\rangle $ and $\cd$ as 
a vector space is the direct sum of $\cg$ and $\tcg$. This ordered triple of algebras $(\cd,\cg$,$\tcg)$ is called the {\bf Manin triple}. 

One can see that dimensions of the subalgebras are equal and that bases 
$(X_i), (\tilde X^i)$ in the subalgebras can be chosen so that
\be \langle X_i,X_j\rangle =0,\  \langle X_i,\tilde X^j\rangle =\langle 
\tilde X^j,X_i\rangle =\delta_i^j,\  \langle \tilde X^i,\tilde X^j\rangle =0;\lbl{brackets}\ee
in the following we always assume that the bases of $\cd$ are of this form.
Due to the ad-invariance of $\langle .,.\rangle $ the algebraic structure of $\cd$ is determined 
by the structure of maximal isotropic subalgebras; in the basis $(X_i), (\tilde X^i)$ 
the Lie product is given by 
\be [X_i,X_j]={f_{ij}}^k X_k,\ [\tilde X^i,\tilde X^j]={\tilde {f^{ij}}_k} \tilde X^k,\ 
[X_i,\tilde X^j]={f_{ki}}^j \tilde X^k +{\tilde {f^{jk}}_i} X_k. \lbl{liebd}\ee

It is clear that to any Manin triple $({\cal D},{\cal G},{\tilde{\cal G}})$   one can construct the 
dual one by interchanging $\cg \leftrightarrow \tcg$, i.e. interchanging the structure coefficients 
$ {f_{ij}}^k \leftrightarrow {\tilde {f^{ij}}_k}$ and such Manin triples give rise to the same Drinfeld double.  
On the other hand, it might be possible to decompose a given Drinfeld double into more than two Manin triples.

The set of all possible decompositions of the Lie algebra ${\cal D}$ of the Drinfeld double $D$ into different Manin triples, i.e.
all possible decompositions of ${\cal D}$ into two maximal isotropic subalgebras is called the {\bf modular space} ${\cal M(D)}$
of the Drinfeld double. 
In general one can find the modular space if one knows the group of automorphisms of the Lie algebra $Aut({\cal D})$ and a complete list of non--isomorphic 
Manin triples $(\cd,\cg_i,\tcg_i)$ generating the Drinfeld double $D$.\footnote{Manin triples $(\cd,\cg,\tcg)$ and $(\cd,\cg',\tcg')$ are isomorphic if and only if exists 
a $\langle.,.\rangle$-preserving automorphism $\phi$ of ${\cal D}$ s. t. $\phi(\cg)=\cg'$, $\phi(\tcg)=\tcg'$.} Such lists were given in \cite{hlasno:doubles}.
The part of modular space ${\cal M(D)}$ corresponding to Manin triples isomorphic to $(\cd,\cg_i,\tcg_i)$ can then be written
\be\label{mdi} {\cal M(D)}_i = \frac{ Aut({\cal D}) \bigcap O({\rm n},{\rm n},\real)}{{\cal H}_i} \ee
 where the pseudoorthogonal group $O({\rm n},{\rm n},\real)$ consists of linear transformations leaving $\langle.,.\rangle$ invariant\footnote{Evidently the group of inner
automorphisms $In(D) \in Aut({\cal D}) \bigcap O({\rm n},{\rm n},\real)$ since $\langle.,.\rangle$ is ad--invariant.}
 and 
${\cal H}_i$ is the subgroup of transformations leaving the isotropic subalgebras $\cg_i,\tcg_i$ invariant. By coset space we mean for concreteness 
the right coset space $[a]={\cal H} a$.
The whole modular space $\cal M(D)$ is the union of ${\cal M(D)}_i$,
\be {\cal M(D)} = \bigcup_{i} {\cal M(D)}_i. \ee

\section{Poisson--Lie T--dual $\sigma$--models and Drinfeld doubles}
\noindent
Starting from a Drinfeld double one can construct the {\bf Poisson--Lie T--dual 
$\sigma$--models} on it. The construction of the models is described in the papers \cite{klse:dna},\cite{kli:pltd}.
The models have target spaces\footnote{Also a generalization to manifolds on which $G$, resp. $\tilde G$ act freely is possible.} in the Lie groups $G$ 
and $\tilde G$ corresponding to the Lie algebras $\cg$, resp. $\tcg$ of a chosen Manin triple, and are defined on $(1+1)$--dimensional Minkowski spacetime $M$ with 
light--cone coordinates $z,\bar{z}$ by the actions
\be  S =  \int dz d\bar{z}  E_{ij}(g) (g^{-1}\partial_- g)^i(g^{-1}\partial_+ g)^j, \; \; \;   
\tilde{S} =  \int dz d\bar{z}  \tilde{E}^{ij}(\tilde{g}) (\tilde{g}^{-1}\partial_- \tilde{g})_i(\tilde{g}^{-1}\partial_+ \tilde{g})_j, \ee
where the coordinates of elements of $\cg,\tcg$ are written in the bases $(X_i)$, $(\tilde X^j)$, e.g. 
$$ g^{-1}\partial_\pm g = (g^{-1}\partial_\pm g)^i X_i  $$
and the (non--symmetric) metrics $E, \tilde{E}$ are 
\be E(g)=(a(g) + E(e)b(g))^{-1}E(e)d(g), \ee
$E(e)$ is a constant matrix
and $a(g),b(g),d(g)$ are submatrices of the adjoint representation of the group $G$ on $\cd$ in the basis $(X_i,\tilde X^j)$
\be Ad(g)^T  =  \left ( \begin{array}{cc} 
  a(g)&0  \\ b(g)&d(g)  \end{array} \right ). \ee
The matrix $\tilde E(\tilde g)$ is constructed analogously  with 
\be Ad(\tilde{g})^T  =  \left ( \begin{array}{cc} 
  \tilde{d}(\tilde{g})& \tilde{b}(\tilde{g})  \\ 0 & \tilde{a}(\tilde{g})  \end{array} \right ), \hskip7mm \tilde E(\tilde e) = E(e)^{-1}. \ee

These metrics satisfy the generalized isometry conditions. These are most easily written after choosing some coordinates $\phi^i$ on $G$, resp. 
$\tilde{\phi}_i$ on $\tilde{G}$ and defining $F_{ij}$, resp. $\tilde{F}^{ij}$ such that after evaluation of $g$ in terms of $\phi^i$s is
$$  E_{ij}(g) (g^{-1}\partial_- g)^i(g^{-1}\partial_+ g)^j = F_{ij}(\phi) \partial \phi^{i} \bar{\partial} \phi^{j},$$
resp.
$$  \tilde{E}^{ij}(\tilde{g}) (\tilde{g}^{-1}\partial_- \tilde{g})_i(\tilde{g}^{-1}\partial_+ \tilde{g})_j = \tilde{F}^{ij}(\tilde{\phi}) 
\partial \tilde{\phi}_{i} \bar{\partial} \tilde{\phi}_{j},$$
and furthermore defining the left--invariant vector fields on the groups $G,\tilde G$
$$ {\bf e}_a = e_{a}^{k} \frac{\partial}{\partial \phi^k}, \; \; {\bf \tilde e}^a = \tilde e^{a}_{k} \frac{\partial}{\partial \tilde \phi_k} $$
corresponding to the basis vectors $X_a,\tilde X^a$ of the Lie algebras $\cg,\tcg$, i.e. satisfying
$$ [{\bf e}_a,{\bf e}_b]= {f_{ab}}^c {\bf e}_c, \; [{\bf \tilde e}^a,{\bf \tilde e}^b]= {\tilde {f^{ab}}_c} {\bf \tilde e}^c. $$
The generalized isometry condition then reads 
\be\label{cond1}
(\lie_{{\bf e}_{a}} F)_{ij}  = {\tilde {f^{bc}}_{a}} F_{ik} e^{k}_{b} F_{lj} e^{l}_{c},  
\ee
where $\lie$ denotes Lie derivative, and similarly for $\tilde{F}$
\be\label{cond2}
(\lie_{{\bf \tilde{e}}^{a}} \tilde{F})^{ij}  = {{{f}}_{bc}}^{a} \tilde{F}^{ik} \tilde{e}_{k}^{b} \tilde{F}^{lj} \tilde{e}_{l}^{c}.  
\ee
In the case of isometry, i.e. non--Abelian duality one has $(\lie_{{\bf e}_{a}} F)_{ij}  =0$, i.e. ${ \tilde {f^{bc}}_{a} }=0$.

The duality can be most straightforwardly understood from the fact that both equations of motion of the above given Lagrangian systems can be reduced from an equation of 
motion on the whole Drinfeld double, $l: \ M \rightarrow D$
\be\label{emkl}
\langle (\partial_{\pm} l) l^{-1}, {\cal E}^{\pm} \rangle =0,
\ee
where subspaces ${\cal E}^{+}= {\rm span}(X_i+E_{ij}(e) \tilde{X}^j)$, ${\cal E}^{-}= {\rm span}(X_i-E_{ji}(e) \tilde{X}^j)$ 
are orthogonal with respect to $\langle . , . \rangle$ and span the whole Lie algebra $\cd$.
One writes $l = g .\tilde{h}, 
\, g \in G,\, \tilde{h} \in \tilde{G}$ (such decomposition of group elements exists at least at the vicinity of the  unit  element, see \cite{Drinfeld}) 
and eliminates $\tilde{h}$ from (\ref{emkl}), respectively $l = \tilde{g} .h, 
\, h \in G,\, \tilde{g} \in \tilde{G}$ and eliminates $h$ from (\ref{emkl}). The resulting equations of motion
for $g$, resp. $\tilde{g}$ are the equations of motion of the corresponding lagrangian systems (see \cite{klse:dna}).

Also the quantum version of Poisson--Lie T--duality has been investigated, e.g. in \cite{sfe:pltd}, \cite{tseytlinrvu}. One of important discoveries is the fact that 
if one wants to have conformally invariant model on $\tilde{G}$ after T--duality transformation, the adjoint representation of the original group $G$ must be 
traceless, similarly as in the non--Abelian duality \cite{agmr}, \cite{ealagyl}. Therefore if one is interested only in conformally invariant models, 
one seems to be forced to consider only Manin triples whose both isotropic subalgebras have traceless adjoint representation because the same applies 
also for the transition from a model on $\tilde G$ to a model on $G$. This requirement is modified by taking Poisson--Lie T--plurality into account.

Since the equations of motion are deduced \footnote{in certain cases only locally} from equation (\ref{emkl})
defined originally on $D$ without any reference to $\cg,\tcg$, the transitions between Manin triples of the same Drinfeld double $D$
lead to other pairs of $\sigma$--models whose equations of motion are also derived from the same equation on $D$ and in this sense all these models are equivalent.
This generalization of the original T--duality was recently named {\bf Poisson--Lie T--plurality} and its properties on quantum level 
were investigated using path integral methods in \cite{unge}. The Drinfeld doubles most interesting from the point of view of \cite{unge} are  those 
possesing decomposition into several Manin triples, each having one isotropic subalgebra with traceless adjoint representation 
and the other one with traceful adjoint representation. Such Drinfeld double structure enables to construct a conformally non--anomalous duality transformation 
between models on subgroups corresponding to the subalgebras with traceful adjoint representation belonging to different decompositions of the Drinfeld double 
into Manin triples. Some of the Drinfeld doubles investigated in this paper,
namely  the 1--parameter classes of Drinfeld doubles with Lie algebras  $sl(2,\real) \oplus sl(2,\real)$ and $so(1,3)$, are of this kind. Therefore, a conformal 
duality transformation connecting different models on the euclidean group $V$ (the traceful component of the corresponding Manin triples) should exist. This 
provides further motivation for investigation of these Drinfeld doubles.

The pairs of $\sigma$--models whose Manin triples are connected by a transformation from the group ${\cal H}$ introduced in (\ref{mdi})
describe the same pair of models in different coordinates 
since the maximal isotropic subalgebras are the same, only their bases have changed.
Note that since we have a transformation of the whole group, one uses the automorphisms of the group $D$, i.e. $Aut(D)$ and this might not coincide 
with $Aut({\cal D})$ for non--simply connected groups.

\noindent {\bf Example:} A known example is the modular space of Abelian 2n--dimensional Drinfeld double $D=\real^{2n}$. In this case all Manin triples are isomorphic, 
$Aut({\cal D})=GL(2n,\real)$,
${\cal H}=GL(n,\real)$ ($X_i'=X_k A^k_i,\ \tilde X^{'j}=(A^{-1})^j_k \tilde X^k$) and consequently 
$$ {\cal M}(\real^{2n}) = \frac{O({\rm n},{\rm n},\real)}{GL(n,\real)}. $$
If one compactifies the Abelian group $\real^{2n}$ to the torus $T^{2n}$, e.g. in the directions of $X_i,\tilde{X}^i$ and chooses the diameters so that 
$x \simeq y \Leftrightarrow \exists z \in {\cal Z}^{2n}: \, x=y+z$, then only elements of $O(n,n,{\cal Z})$, i.e. matrices with integer entries preserving $\langle.,.\rangle$,
 are automorphisms of the double as a Lie group since the points of the lattice of integers ${\cal Z}^{2n}$ must be mapped back to ${\cal Z}^{2n}$. Similarly ${\cal H} \simeq
GL(n,{\cal Z})$ and  
the modular space of $D=T^{2n}$ is 
$$ \frac{O(n,n,{\cal Z})}{GL(n,{\cal Z})}.$$

In the following we consider only the algebraic structure, 
the Drinfeld doubles as  Lie groups can be obtained in principle by means of exponential map. 
All results can be transferred immediately to connected and simply connected Lie groups, the modular spaces of non--simply connected versions of 
Drinfeld doubles might be different as in the example above. 

\section{Automorphisms of $sl(2,\real)$ and $so(1,3)$}
In this section we briefly review the automorphisms of $sl(2,\compl)$,  $sl(2,\real)$ and $so(1,3)$. The knowledge of these will enable us to construct 
the groups of automorphisms and their subgroups ${\cal H}$.
We use the usual realization of the group $SL(2,\compl)$:
$$ g = \left( \ba{cc} a & b \\ c & d  \ea \right), \; ad - bc =1 $$
and consider the basis of $sl(2,\compl)$
$$ H =  \left( \ba{cc} 1 & 0 \\ 0 & -1  \ea \right), \; E_{+} =  \left( \ba{cc} 0 & 1 \\ 0 & 0  \ea \right), 
\; E_{-} =  \left( \ba{cc} 0 & 0 \\ 1 & 0  \ea \right), $$
with commutation relations 
\be\label{komrel}
[H,E_{+}] = 2 E_{+}, \, [H,E_{-}] = - 2 E_{-}, \, [E_{+},E_{-}] = H.\ee
It is known that all automorphisms of $sl(2,\compl)$ are inner (see e.g. \cite{helgason}), i.e. the map $Ad: \, SL(2,\compl) 
\rightarrow {\rm Aut}(sl(2,\compl)):Ad(g) X= g X g^{-1}$ is onto. In the basis $(X_i)=(H,E_{+},E_{-})$ we may write the matrix\footnote{
The notation is chosen to allow simple expressions later, because of it we have 
$Ad(g \tilde{g})_{ij}=Ad(\tilde{g})_{ik} Ad(g)_{kj}=(Ad(\tilde{g}) Ad(g))_{ij}$.}
 of $Ad(g)$, 
$Ad(g) X_i = Ad(g)_{ij} X_j$:
$$ Ad \left( \left( \ba{cc} a & b \\ c & d  \ea \right) \right) = 
\left( \ba{rrr} a d + b c & -2 a b & 2 c d  \\  - a c & a^2 & -c^2 \\ b d  &  -b^2 & d^2  
\ea \right), \; ad - bc =1. $$

The map $Ad$ is not $1-1$, since 
${\rm Ker} \, Ad = \{ {\bf 1},-{\bf 1} \}$ (after denoting the group unit by ${\bf 1}$). Therefore we have 
\be
{\rm Aut}(sl(2,\compl)) \simeq SL(2,\compl) / {\rm Ker} \, Ad \simeq SL(2,\compl) / \{ {\bf 1},-{\bf 1} \}. \ee 

Concerning the group of automorphisms of $sl(2,\real)$, it is clear that it can be considered as a $3 \times3$ matrix group consisting 
of elements of ${\rm Aut}(sl(2,\compl))$ with only real entries (since after complexification we must have an automorphism of 
${\rm Aut}(sl(2,\compl))$).
This in turn leads to condition on parameters $a,b,c,d$: they must be all real or 
all purely imaginary. 
The matrices with real parameters form the group of inner automorphisms $SL(2,\real) / \{ {\bf 1},-{\bf 1} \}$.
The matrices with all parameters purely imaginary correspond to outer automorphisms and can be obtained by multiplication of the inner
ones by the matrix 
$$S = \left( \ba{rrr} 1 & 0 & 0  \\ 0 & -1 & 0 \\ 0 & 0 & -1  \ea \right)$$
(either from left or right, the choice of order amounts to $a \rightarrow -a, \ d \rightarrow -d$). Therefore the group of automorphisms 
of $sl(2,\real)$ is
\be
{\rm Aut}(sl(2,\real)) = \{ {\bf 1},S \} \triangleright (SL(2,\real) / \{ {\bf 1},-{\bf 1} \} )   
\ee
where $\triangleright$ denotes semidirect product.

The automorphisms of $so(1,3)$ are known (see e.g. \cite{gelfand}) to be either inner (in $1-1$ correspondence with elements of $SO(1,3)_{+}$) or correspond to 
composition of an element of $SO(1,3)_{+}$ and the space inversion $P$.

\section{Modular spaces of Drinfeld doubles with the Lie algebra $sl(2,\real) \oplus sl(2,\real)$}

It is known (see \cite{hlasno:doubles}) that there are two classes of non--isomorphic Drinfeld doubles with the Lie algebra $sl(2,\real) \oplus sl(2,\real)$.
\footnote{The subalgebras are denoted and their bases are chosen according to the Bianchi classification, in the common notation 
${\bf 9} \equiv su(2)$, ${\bf 8} \equiv sl(2,\real)$,
${\bf V}$ is the euclidean algebra, ${\bf 7_0}$, ${\bf 6_0}$, ${\bf 7_a}$ and ${\bf 6_a}$ are certain solvable Lie algebras, commutation relations of each of 
these algebras are contained in definitions of Manin triples below.}

\begin{enumerate}
\item 2--parameter class of Drinfeld doubles whose bilinear invariant form is $\langle . , . \rangle = \frac{a}{4b(a-1)^2} K^{1} - 
\frac{a}{4b(1+a)^2} K^{2} $, where $K^{1},K^{2}$ are Killing forms of simple components $sl(2,\real)^1$,$sl(2,\real)^2$,
$a > 1$  and 
the parameter $b \in { \real} - \{ 0 \}$ corresponds to rescaling of $\langle .  , . \rangle$. This Drinfeld double can be decomposed 
only into 
Manin triples isomorphic to
$${\bf ( 6_{a}|6_{1/a}.i|b)} : \;
  [X_1,X_2]=-a X_2-X_3, \ [X_2,X_3] = 0, \ [X_3,X_1] = X_2+ a X_3, $$
$$ [\tilde{X}^1,\tilde{X}^2]=- b ( \frac{1}{a} \tilde{X}^2+\tilde{X}^3), \, 
[\tilde{X}^2,\tilde{X}^3] = 0, \; 
[\tilde{X}^3,\tilde{X}^1] = b (\tilde{X}^2+ \frac{1}{a} \tilde{X}^3) $$ 
and its dual.
\item 1--parameter class of Drinfeld doubles whose bilinear invariant form is $\langle . , . \rangle = \frac{1}{4b}K^{1} -\frac{1}{4b} 
K^{2}$, the parameter $b \in {\real}^{+}$ (i.e. $b >0$) corresponds to rescaling of 
$\langle . , . \rangle$. Any such Drinfeld double possesses decompositions into four non--isomorphic Manin triples, namely 
$${\bf ( 8|5.i|b) } : \;  [X_1,X_2]=-X_3, \ [X_2,X_3] = X_1, \ [X_3,X_1] = X_2, $$
$$ [\tilde{X}^1,\tilde{X}^2]=- b \tilde{X}^2, \ 
[\tilde{X}^2,\tilde{X}^3] = 0 , \ [\tilde{X}^3,\tilde{X}^1] = b \tilde{X}^3 $$ 
and 
$${\bf ( 6_0|5.iii|b) } : \;
[X_1,X_2]=0, \ [X_2,X_3] = X_1, \ [X_3,X_1] = -  X_2, $$
$$[\tilde{X}^1,\tilde{X}^2]=  0 , \ 
[\tilde{X}^2,\tilde{X}^3] = - b\tilde{X}^2, \
[\tilde{X}^3,\tilde{X}^1] =  b\tilde{X}^1 $$
and their duals.
\end{enumerate}

In order to identify all possibilities how a given Manin triple can  be contained in the given Drinfeld double, i.e. in how many ways 
one may decompose  
the Lie algebra of the given Drinfeld double into a pair of maximal isotropic subalgebras $\cg'$, $\tcg'$ isomorphic as Lie algebras 
to given $\cg,\tcg$
we firstly realize that the doubles considered have 2 significant subalgebras, namely the simple components $sl(2,\real)^1$,
$sl(2,\real)^2$. 
These of course don't depend on the choice of $\cg$, $\tcg$. Therefore we firstly find a transformation $T$ from the original 
dual basis of the 
Manin triple $(\cd,\cg,\tcg)$ $(X(j))_{j=1\ldots 6} = (X_1, \ldots, \tilde{X}^3)$ to the bases of $sl(2,\real)^1$,$sl(2,\real)^2$ 
$(Y(j))_{j=1\ldots 6} = 
(H^1,E_{+}^1,E_{-}^1,H^2,E_{+}^2,E_{-}^2)$ satisfying the usual commutation relations (\ref{komrel}) in $sl(2,\real)^1$, resp. 
$sl(2,\real)^2$, 
$$Y(j)=T_{jk} X(k).$$
Of course, the bases of $sl(2,\real)$ are defined only up to automorphisms of $sl(2,\real)$, so we just fix one possible $T$. Applying the same 
transformation $T$ 
on any dual bases of any possible maximal isotropic subalgebras $\cg'$, $\tcg'$ we find again bases of $sl(2,\real)^1$, resp. 
$sl(2,\real)^2$ satisfying 
the same commutation relations. It follows that any decomposition of $\cd$ into $\cg'$, $\tcg'$ can be obtained by application of 
the map $T^{-1} A T$ on $\cg,\tcg$ where $A$ is an automorphism of 
$sl(2,\real) \oplus sl(2,\real)$ preserving the simple subalgebras, i.e. it has in the basis $Y(j)$  
the matrix
$$ A= \left( \ba{rrrrrr} 
 \alpha  \delta +\beta   \gamma   &  -2 \alpha  \beta   &  2 \gamma   \delta  &  0 &  0 &  0 \\
 -\alpha  \gamma   &  \alpha ^2 &  -\gamma  ^2 &  0 &  0 &  0 \\
 \beta   \delta  &  -\beta  ^2 &  \delta ^2 &  0 &  0 &  0 \\
 0 &  0 &  0 &  \kappa  \nu +\lambda  \mu &  -2 \kappa  \lambda  &  2 \mu \nu  \\
 0 &  0 &  0 &  -\kappa  \mu &  \kappa ^2 &  -\mu^2 \\
 0 &  0 &  0 &  \lambda  \nu  &  -\lambda ^2 &  \nu ^2
\ea \right), \, \alpha \delta - \beta \gamma = \kappa \nu - \lambda \mu =1 $$
(The transformation $\tilde{T}^{-1} \Omega A \tilde{T}$ where $\Omega$ interchanges $sl(2,\real)^1$ and  $sl(2,\real)^2$  
is not allowed since it changes the bilinear form $\langle . , . \rangle$.)
The set of all pairs of maximal isotropic 
subalgebras $\cg'$, $\tcg'$ 
coincides with ${\rm Aut} ( sl(2,\real)) \oplus {\rm Aut}(sl(2,\real) ) / {\cal H}$.
In the following subsections we investigate the subgroups ${\cal H}$ for different Drinfeld doubles and Manin triples. We don't consider 
separately the Manin triple and its dual, because there is a $1-1$ correspondence between them, one just interchanges 
$\cg \leftrightarrow \tcg$.

\subsection{The 2--parameter class}
In this case 
$$ T_{( 6_{a}|6_{1/a}.i|b)} = \left( \ba{rrrrrr} 
\frac{1}{a-1} &  0 &  0 &  \frac{a}{b(a-1)} &  0 &  0 \\
 0 &  0 &  0 &  0 &  -1 &  1 \\
 0 &  -\frac{a}{2b (a-1)^2} &  \frac{a}{2b(a-1)^2} &  0 &  0 &  0 \\
 -\frac{1}{a+1} &  0 &  0 &  \frac{a}{b(a+1)} &  0 &  0 \\ 0 &  1 &  1 &  0 &  0 &  0 \\
 0 &  0 &  0 &  0 &  -\frac{a}{2b(a+1)^2} &  -\frac{a}{2b(a+1)^2}
\ea \right). $$
In order to find ${\cal H}$ we compute $T_{( 6_{a}|6_{1/a}.i|b)}^{-1} A 
T_{( 6_{a}|6_{1/a}.i|b)}$ and impose the condition that it is block diagonal, i.e. leaves the isotropic 
subalgebras $\cg,\tcg$ invariant. We find that $ \beta=\gamma=\lambda=\mu=0, \delta = 1/\alpha, \nu = 1/\kappa$; $\alpha,\kappa$ might 
be real (inner automorphisms) or purely imaginary (outer automorphisms), their signs are irrelevant. 
Therefore we have ${\cal H}= ( \{ {\bf 1},S \} \times {\real}^{+})  \times (\{ {\bf 1},S \} \times {\real}^{+} )$ and finally 
we find that the modular space ${\cal M(D)}$ consists of two components (corresponding to Manin triples with $\cg= 6_{a}$ 
and $\tcg=6_{1/a}$ resp. $\cg=6_{1/a}$ and $\tcg=6_{a}$), each isomorphic  to the homogeneous space
$$ {\cal M(D)}_{( 6_{a}|6_{1/a}.i|b)} \simeq
\frac{ \{ {\bf 1},S \} \triangleright (SL(2,\real) / \{ {\bf 1},-{\bf 1} \}  ) \times (\{ {\bf 1},S \} \triangleright SL(2,\real) / \{ {\bf 1},-{\bf 1} \} ) }{   
(\{ {\bf 1},S \} \times \real^{+} ) \times ( \{ {\bf 1},S \} \times \real^{+}  )} \simeq$$ 
$$\frac{SL(2,\real)}{ \real - \{ 0 \} }  
\times \frac{SL(2,\real)}{ \real - \{ 0 \} }. $$

\subsection{The 1--parameter class}
Firstly we study the possible decompositions of the Drinfeld double into the Manin triple ${\bf ( 8|5.i|b) }$. The matrix $T$ is
$$ T_{ ( 8|5.i|b) } =
\left( \ba{rrrrrr} 
1 &  0 &  0 &  \frac{1}{b} &  0 &  0 \\
 0 &  1 &  -1 &  0 &  \frac{1}{2b} &  \frac{1}{2b} \\
 0 &  0 &  0 &  0 &  \frac{1}{2b} &  -\frac{1}{2b} \\
 -1 &  0 &  0 &  \frac{1}{b} &  0 &  0 \\
 0 &  1 &  1 &  0 &  -\frac{1}{2b} &  \frac{1}{2b} \\
 0 &  0 &  0 &  0 &  -\frac{1}{2b} &  -\frac{1}{2b}
 \ea \right).$$
By the same argument as above we find that ${\cal H}$ consists of matrices $T_{ ( 8|5.i|b) }^{-1} A T_{ ( 8|5.i|b) }$ s. t. 
$ \beta=\gamma=\lambda=\mu=0, \delta = 1/\alpha, \nu = 1/\kappa, \kappa^2= 1/ \alpha^2$. Imposing these conditions the matrices 
$T_{ ( 8|5.i|b) }^{-1} A T_{ ( 8|5.i|b) }$  depend only on $\alpha^2$, $\alpha$ may be real or purely imaginary, i.e.
${\cal H} \simeq ( \{ {\bf 1},S \}  \times {\real}^{+}  ) $ and we obtain that this part of modular space is isomorphic 
to 
$$ {\cal M(D)}_{ (8|5.i|b) } \simeq \frac{ ( \{ {\bf 1},S \} \triangleright (SL(2,\real) / \{ {\bf 1},-{\bf 1} \}  ) 
\times ( \{ {\bf 1},S \} \triangleright (SL(2,\real) / \{ {\bf 1},-{\bf 1} \}  )  }{   ( \{ {\bf 1},S \} \times {\real}^{+} )  }. $$ 
In the whole modular space ${\cal M(D)}$ it will again appear twice because of possible interchange of isotropic subalgebras.

Secondly we study the possible decompositions of the Drinfeld double into the Manin triple ${\bf ( 6_0|5.iii|b)  }$. The matrix $T$ is
$$ T_{ ( 6_0|5.iii|b) } = \left( \ba{rrrrrr}
0 &  0 &  1 &  0 &  0 &  \frac{1}{b} \\
 0 &  0 &  0 &  1 &  1 &  0 \\
 \frac{1}{2b} &  \frac{1}{2b} &  0 &  0 &  0 &  0 \\
 0 &  0 &  1 &  0 &  0 &  -\frac{1}{b} \\
 -1 &  1 &  0 &  0 &  0 &  0 \\
 0 &  0 &  0 &  \frac{1}{2b} &  -\frac{1}{2b} &  0
\ea \right), $$
${\cal H}$ consists of transformations with $ \beta=\gamma=\lambda=\mu=0, \delta = 1/ \alpha, \nu = 1/ \kappa$, $\alpha,\kappa$ again might 
be real or purely imaginary, i.e.  
${\cal H}=  ( \{ {\bf 1},S \} \times {\real}^{+})  \times (\{ {\bf 1},S \} \times {\real}^{+} ) )$ and
this part of modular space is, as in the 2--parameter class,  isomorphic to 
$$ {\cal M(D)}_{ ( 6_0|5.iii|b) } \simeq \frac{SL(2,\real)}{ \real - \{ 0 \} }  
\times \frac{SL(2,\real)}{ \real - \{ 0 \} }. $$

The whole modular space then consists of four pieces, two isomorphic to 
${\cal M(D)}_{ ( 8|5.i|b) } $ and two to ${\cal M(D)}_{ ( 6_0|5.iii|b) } $.

\section{Modular spaces of Drinfeld doubles with the Lie algebra $so(1,3)$}

It is known (see \cite{hlasno:doubles}) that there are two classes of non--isomorphic Drinfeld doubles with the Lie algebra $so(1,3)$.
For our considerations it seems appropriate to consider also its complexification 
$so(1,3)_{\compl}=sl(2,\compl)^1 \oplus sl(2,\compl)^2$. Then we have
\begin{enumerate}
\item 2--parameter class of Drinfeld doubles whose bilinear invariant form is in complexification $\langle . , . \rangle = 
\frac{ia}{4b(a-i)^2} K^{1}-\frac{ia}{4b(i+a)^2}  K^{2}$, where $K^{1},K^{2}$ are Killing forms of simple components,
$a \geq 1$  and the parameter $b \in { \real} - \{ 0 \}$ corresponds to rescaling of $\langle . , . \rangle$. This Drinfeld double can be decomposed 
only into Manin triples isomorphic to
$$ {\bf (7_{a}|7_{ \frac{1}{a} }|b) }: \; [X_1,X_2]=-a X_2+X_3, \; [X_2,X_3] = 0, \; [X_3,X_1] = X_2+ a X_3, 
 $$
$$ [\tilde{X}^1,\tilde{X}^2]= b ( -  \frac{1}{a} \tilde{X}^2 + \tilde{X}^3), \, 
[\tilde{X}^2,\tilde{X}^3] = 0, \; 
[\tilde{X}^3,\tilde{X}^1] = b (\tilde{X}^2+ \frac{1}{a} \tilde{X}^3),  $$ 
and its dual.
\item 1--parameter class of Drinfeld doubles whose bilinear invariant form is in complexification $\langle . , . \rangle = 
\frac{i}{4b}K^{1} -\frac{i}{4b} K^{2}$, the parameter $b \in {\real}^{+}$  corresponds to rescaling of 
$\langle . , . \rangle$. Any such Drinfeld double possesses decompositions into six non--isomorphic Manin triples, namely 
$${\bf (9|5|b )} : \;  [X_1,X_2]=X_3, \; [X_2,X_3] = X_1, \; [X_3,X_1] = X_2, $$
  $$  [\tilde{X}^1,\tilde{X}^2]=- b \tilde{X}^2, \, 
[\tilde{X}^2,\tilde{X}^3] = 0 , \; 
[\tilde{X}^3,\tilde{X}^1] = b \tilde{X}^3,  $$ 
$$ {\bf ( 8|5.ii|b) }  : \;  [X_1,X_2]=-X_3, \; [X_2,X_3] = X_1, \; [X_3,X_1] = X_2, $$
$$ [\tilde{X}^1,\tilde{X}^2]=0, \, [\tilde{X}^2,\tilde{X}^3] = b  \tilde{X}^2 , \; 
[\tilde{X}^3,\tilde{X}^1] = - b \tilde{X}^1,  $$ 
and 
$$ {\bf ( 7_0|5.ii|b) }  : \;  [X_1,X_2]=0, \; [X_2,X_3] = X_1, \; [X_3,X_1] = X_2, $$
$$  [\tilde{X}^1,\tilde{X}^2]=  0 , \, [\tilde{X}^2,\tilde{X}^3] = b \tilde{X}^2, \; 
[\tilde{X}^3,\tilde{X}^1] = -b \tilde{X}^1, $$
and their duals.
\end{enumerate}

Concerning the group of automorphisms of these Drinfeld doubles, one may easily check that the space inversion $P$, which in $so(1,3)_{\compl}$
interchanges the bases of $sl(2,\compl)^1$ and $sl(2,\compl)^2$, i.e. $P(Y(j)) = Y(j \pm 3)$, in both cases changes 
$\langle . , . \rangle$\footnote{In the 1--parameter class $P$ reverses the sign of $\langle . , . \rangle$, 
in the 2--parameter class the change is more complicated.} and only inner automorphism remain. In order to find the subgroups leaving the Manin triples
invariant we proceed similarly to the case $sl(2,\real) \oplus sl(2,\real)$; the difference is  that we have to use the complexification of 
 $so(1,3)$, perform all computations, and at the end restrict the possible transformations to those with only real entries.
In this way we find
\begin{enumerate}
\item in the case of the 2--parameter class the transformation to the bases of  $so(1,3)_{\compl}=sl(2,\compl)^1 \oplus sl(2,\compl)^2$ is
$$ T_{(7_{a}|7_{\frac{1}{a}}|b)} = \left( \ba{rrrrrr} \frac{1}{a+i} & 0 & 0 & -\frac{ia}{b(a+i)} & 0 & 0 \\
 0 & 0 & 0 & 0 & i & 1 \\
 0 & -\frac{a}{2b(a+i)^2} & -\frac{ia}{2b(a+i)^2} & 0 & 0 & 0 \\
 -\frac{1}{i-a} & 0 & 0 & -\frac{ia}{b(i-a)} & 0 & 0 \\
 0 & 0 & 0 & 0 & -i & 1 \\
 0 & -\frac{a}{2b(i-a)^2} & \frac{ia}{2b(i-a)^2} & 0 & 0 & 0 \ea \right). $$
As before we look for the subgroup ${\cal H}$ of transformations leaving the Manin triple invariant and find
$ \beta=\gamma=\lambda=\mu=0, \delta = 1/\alpha, \nu = 1/\kappa, $ together with reality conditions
$Im(\alpha^2+\kappa^2)=0, Re(\alpha^2-\kappa^2)=0$; the elements of $T^{-1} A T$ depend only on $\alpha^2$ and $\kappa^2$. Therefore
one may introduce two real parameters $\psi \in \real^{+}, \phi \in \langle 0,2 \pi )$ and  write $\alpha^2=\psi \exp(i\phi), \kappa^2=\psi \exp(-i\phi)$ and identify the subgroup ${\cal H}$ with
the commutative group of rotations and boosts along the same axis ${\cal H}= U(1) \times \real^{+}$. The modular space 
${\cal M(D)}$ is equal to 
$${\cal M(D)}_{(7_{a}|7_{\frac{1}{a}}|b)} \simeq \frac{SO(1,3)_{+}}{{\cal H}} \simeq \frac{SO(1,3)_{+}}{U(1) \times \real^{+}}$$ 
if $a=1$ (the self--dual case $\cg \simeq \tcg$)
and consists of two such components if $a >1$.
\item in the case of the 1--parameter class the transformations to the bases of  $so(1,3)_{\compl}=sl(2,\compl)^1 \oplus sl(2,\compl)^2$ are respectively
$$T_{ (9|5|b )}  = \left( \ba{rrrrrr}
-i & 0 & 0 & -\frac{1}{b} & 0 & 0 \\
 0 & 0 & 0 & 0 & i & 1 \\
 0 & \frac{1}{2b} & \frac{i}{2b} & 0 & -\frac{i}{4b^2} & \frac{1}{4b^2} \\
 i & 0 & 0 & -\frac{1}{b} & 0 & 0 \\
 0 & 0 & 0 & 0 & 1 & i \\
 0 & -\frac{i}{2b} & -\frac{1}{2b} & 0 & \frac{1}{4b^2} & -\frac{i}{4b^2}
\ea \right),$$
$$T_{ ( 8|5.ii|b) }  = \left( \ba{rrrrrr}
0 & 0 & i & 0 & 0 & \frac{1}{b} \\
 -2ib & 2b & 0 & 1 & i & 0 \\
 0 & 0 & 0 & -\frac{1}{4b^2} & \frac{i}{4b^2} & 0 \\
 0 & 0 & i & 0 & 0 & -\frac{1}{b} \\
 0 & 0 & 0 & 1 & i & 0 \\
 -\frac{i}{2b} & -\frac{1}{2b} & 0 & -\frac{1}{4b^2} & \frac{i}{4b^2} & 0
\ea \right),$$
$$T_{ ( 7_0|5.ii|b) }  = \left( \ba{rrrrrr}
0 & 0 & i & 0 & 0 & \frac{1}{b} \\
 1 & i & 0 & 0 & 0 & 0 \\
 0 & 0 & 0 & \frac{i}{2b} & \frac{1}{2b} & 0 \\
 0 & 0 & -i & 0 & 0 & \frac{1}{b} \\
 i & 1 & 0 & 0 & 0 & 0 \\
 0 & 0 & 0 & -\frac{1}{2b} & -\frac{i}{2b} & 0
\ea \right).$$
The subgroup ${\cal H}$ is determined for the Manin triples ${\bf  (9|5|b) }$ and ${\bf (8|5.ii|b) }$ by the conditions 
$ \beta=\gamma=\lambda=\mu=0, \delta = 1/\alpha, \nu = 1/\kappa, \kappa^2=1/\alpha^2$ together with reality conditions
\be\label{reality}
Im(\alpha^2+\kappa^2)=0, \; Re(\alpha^2-\kappa^2)=0 \ee
 and the elements of $T^{-1} A T$ depend only on $\alpha^2$ and $\kappa^2$. 
Using $\kappa^2=1/\alpha^2$ the reality conditions (\ref{reality}) can be written 
$$Im \alpha^2 = \frac{1}{|\alpha|^4} Im \alpha^2, \; Re \alpha^2 = \frac{1}{|\alpha|^4} Re \alpha^2,$$
and we find that there is only one free parameter $\alpha^2 \in \compl$ such that. $|\alpha^2|=1$. Such subgroup ${\cal H}$ is evidently isomorphic to $U(1)$.
 Concerning the Manin triple ${\bf ( 7_0|5.ii|b) }$, ${\cal H}$ is in this case given by the same conditions as in the 2--parameter class, i.e.
${\cal H}= U(1) \times \real^{+}$.

The whole modular space then consists of six pieces, four ($\Leftarrow$ Manin triples $ (9|5|b )$,$( 8|5.ii|b) $ and their duals) of them are isomorphic to 
$${\cal M(D)}_{(9|5|b )} \simeq {\cal M(D)}_{ (8|5.ii|b)} \simeq  \frac{SO(1,3)_{+}}{U(1)} $$
and two ($\Leftarrow$ Manin triple $( 7_0|5.ii|b)$ and its dual) are isomorphic to 
$${\cal M(D)}_{( 7_0|5.ii|b)} \simeq \frac{SO(1,3)_{+}}{U(1) \times \real^{+}}.$$

If one tries to better understand the structure of ${\cal M(D)}$ in this case, one discovers that
the Manin triple ${( 7_0|5.ii|b)}$ can be viewed as a contraction of ${( 8|5.ii|b)}$. If one puts  
$Z_{1,2}=\epsilon X_{1,2},Z_{3}= X_{3}, \tilde{Z}^{1,2}= \tilde{X}^{1,2}/\epsilon,\tilde{Z}^{3}= \tilde X^{3}$, then 
for $\epsilon>0$ $(Z_i),(\tilde{Z}^i)$ form another pair of dual bases of ${ ( 8|5.ii|b) }$ and for $\epsilon=0$ one obtains the commutation relations
of ${( 7_0|5.ii|b)}$. One may be therefore tempted to consider ${( 7_0|5.ii|b})$ as some limiting case of ${( 8|5.ii|b)}$ and 
the corresponding part of the modular space ${\cal M(D)}_{( 7_0|5.ii|b)}$ as a closure of compactification of ${\cal M(D)}_{( 8|5.ii|b)}$.
Further support for this interpretation comes from the fact that after fixing a transformation $C: \, {\bf {( 7_0|5.ii|b)}} \rightarrow {\bf ( 8|5.ii|b)}$ 
one may associate unambiguously to each Manin triple of type ${( 7_0|5.ii|b)}$ a one--parameter family of Manin triples isomorphic to ${ ( 8|5.ii|b) }$
that are obtained by transformations $C T_{( 7_0|5.ii|b)}^{-1} A T_{( 7_0|5.ii|b)}$ where $A ={\rm diag}(1,\psi,1/\psi,1,\psi,1/\psi) \in {\cal H}_{( 7_0|5.ii|b)}$.

The problem is that this $1-1$ correspondence between Manin triples of type ${( 7_0|5.ii|b)}$ and one--parameter families of Manin triples 
${( 8|5.ii|b)}$ is unique only after fixing of $C$ and $T_{( 7_0|5.ii|b)}$; after their change the relationship changes accordingly.
Therefore, it is not an intrinsic geometric property of the modular space. Also if the ``limit conjecture'' is true,
the transformation $(T_{{( 7_0|5.ii|b)}})^{-1} T_{( 8|5.ii|b)}$ 
should be in some sense a limit of transformations $T_{( 8|5.ii|b)}^{-1} A T_{( 8|5.ii|b)}$ taking 
${( 8|5.ii|b)}$ to isomorphic Manin triples. One can easily check that it is not possible, since 
coinciding elements in $T_{( 8|5.ii|b)}^{-1} A T_{( 8|5.ii|b)}$  are not equal in $(T_{{( 7_0|5.ii|b)}})^{-1} T_{( 8|5.ii|b)}$.
Therefore the interpretation of the structure of the modular space ${\cal M(D)}$ suggested above seems unjustified at the present moment and
the parts of the modular spaces corresponding to non--isomorphic Manin triples might be disconnected.

Similar contraction can be found also for Manin triples ${( 7_0|5.ii|b)}$ and ${ ( 9|5|b) }$, resp. $( 6_0|5.iii|b)$ and ${ ( 8|5.i|b) }$ after suitable 
choice of bases, but the same problem of non-existing limit of $T_{(\ldots)}^{-1} A T_{(\ldots)}$ etc. arises. 

\end{enumerate}

\section{Conclusions}

We have presented four examples of modular spaces of Drinfeld doubles. These are rather different from the known Abelian one, mainly $Aut({\cal D}) \bigcap O(d,d,\real)$ 
is in these cases (almost) the group of inner automorphisms $In({\cal D})$, whereas in the Abelian case $In({\cal D})=\{ {\bf 1} \}$. In this sense they represent 
other extremal cases of modular spaces. Also we have encountered the fact that the modular spaces might be composed of parts of different dimensions, as in the both 
1--parametric classes. Consequently, after fixing one concrete T--plurality transformation and applying it to (pairs of) models on some set of isomorphic Manin triples 
one may obtain just one (pair of) models on another Manin triple written in different coordinates and vice versa. 

The method we have used can be applied also to non--semisimple Drinfeld doubles, especially if they contain some distinct subalgebras invariant with respect 
to automorphisms (e.g. 
the double $SL(2,\real) \times \real^3$) and one can therefore easily find the group of automorphisms. On the other hand there are Drinfeld doubles (e.g. two classes
of doubles with ${\cal D}=sl(2) \triangleright \real^3$ or some solvable ones) arising from quite a large number of non--isomorphic Manin triples and 
the structure of modular spaces would be in these cases even more complicated. 

We should also mention again that we have assumed that the Drinfeld double is simply connected. Therefore all automorphisms of Lie algebra could be raised
to automorphisms of Lie group. This may not be true for not simply--connected doubles. In the semisimple 
cases this is probably not significant, e.g. the center ${\cal Z}(SL(2,\compl)_{\real})= \{ {\bf 1}, {\bf -1} \}$  and therefore automorphisms 
of $SL(2,\compl)_{\real}$ can be factored to automorphisms of the only other connected Lie group with the same Lie algebra 
$SO(1,3)_{+}$ since $\phi({\cal Z})={\cal Z}$ and consequently $\phi({\bf -1})={\bf -1}$; the center of ${\cal Z}(SL(2,\real) \times SL(2,\real))$ consists of four 
elements and some moderate changes in the modular spaces may occur depending on the choice of the connected group.
One may also assume that if one considers dualities in quantum theories, only discrete subsets of the presented 
modular spaces might be relevant.
\smallskip

\noindent {\bf Acknowledgment:} I thank professor Ladislav  Hlavat\'y for discussions, encouragement and help in preparation of this paper and also 
Rikard von Unge for conversations and comments.

\end{document}